# Density dependent equations of state for metal, nonmetal, and transition states for compressed mercury fluid

*M.H. Ghatee*[*], *M. Bahadori*

(Department of Chemistry, Shiraz University, Shiraz, 71454, Iran)

Fax: +98 711 228 6008

E-mail: ghatee@sun01.susc.ac.ir

[*] Corresponding Author

# Density dependent equations of state for metal, nonmetal, and transition states for compressed mercury fluid

Abstract

Analytical equations of state are presented for fluid mercury in metal, nonmetal, and in metal-nonmetal transition states. Equations of state for metal and nonmetal states are simple in form but the complexities of transition state leads to a complex fourth-order equation. The interatomic potential function used to describe the metal state have a hard repulsive wall, and that of nonmetal state is the same as potential function of non-polar fluid with induced dipole intermolecular interaction. Metal-nonmetal transition occurs in the liquid density range $11-8\ g/cm^3$, and a density dependent interaction potential which gradually changes from a pure metal interaction to a nonmetal interaction, on going from metal state to nonmetal state in the transition region, is used. Well-depth and the position of potential minimum are presented as temperature dependent quantities; their calculated values for the metal state are typically within 5.0% and 0.33% of the experimental value, respectively. The calculated well-depth for nonmetal state is smaller than the experimental value indicating the effect of high pressure $P\rho T$ data used, which pushes a pair of mercury atom further together into the repulsive side. In the transition region, calculated well-depths are 2-3 order of magnitudes larger than for the metal state, and contain a sharp rising edge and a steep falling having a singularity characteristic of phase transition.

## 1. Introduction

Mercury has the lowest critical temperature of any fluid metals. Therefore, it has been investigated experimentally with accurate measurements in the critical region. These measurements involve its magnetic, electrical, structural, optical and thermophysical properties with optimal control of temperature in the critical region.[1] The fundamental difficulty in dealing with fluid metals is that the electronic structures of liquid and gas phases are completely different. The electronic structure of mercury (and other molten metals) at low temperatures can be well approximated by the structure of the solid metals, and thus the thermodynamic properties of metals can be obtained by the same cohesion mechanism of the solid metals. Experimental data clearly represent that mercury near its triple point (at densities larger than 11 g/cm$^3$) follows the nearly free electron theory of metals that considers the nuclei completely shielded by delocalized electrons.[2] At lower densities (9 to 8 g/cm$^3$), the cohesion mechanism of its atoms will be suppressed by a partial localization of electrons and the metallic character is changed to a nonmetallic kind and a gradual metal non-metal transition (M-NMT) occurs. Recently Nagel et al.[3] have calculated the structure factors of expanded liquid mercury for a wide range of density using the effective pair potential obtained from pseudopotential perturbation theory for liquid metal and by Lennard-Jones potential for corresponding vapor. Although these pair potential are valid in the two corresponding limiting cases, their validity are lost for those states that are near M-NMT.[3,4]

In addition, experimental investigations of monovalent liquid cesium and rubidium indicate a gradual metal non-metal transition near the critical region but the properties of this transition for a divalent metal are completely different. In this state, fractions of alkali metals atoms form chemically bonded dimers, however, mercury because of its 6s$^2$ closed shell ground electronic state does not undergo such a process. From experimental data, it was concluded that metal-nonmetal transition in liquid mercury is mainly due to lack of overlapping between the 6s and 6p bands.[5,6,7]

For mercury, data of precise measurements of electrical conductivity,[8] Hall coefficient,[9] NMR studies,[10] sound velocity,[11,12] and equation of state,[13] are available. These data clearly demonstrate that a radical change in the atomic states occur from phase to phase. For instance far below the critical point, the liquid is highly conducting but the exiting vapor does not. More close to the liquid-vapor critical point, where the distinction between the two existing phases vanishes, the electrical conductivity, Knight shift, and optical properties show that the non-metallic behavior is present in both phases.

The major reason for the prediction of thermodynamic properties of mercury and other metals lie on the fact that the intermolecular interaction highly depends on temperature and density. Density dependence of mercury potential function has been subjected to the critical investigations,[12] by using the experimental structure factor and theoretical modeling.

Near the critical point especially in M-NMT region, there is no reliable theoretical method to derive effective potential function for liquid metals accurately. Indeed, at high densities near the triple point the effective pair potential function, obtained from nearly free electron model, can reproduces the thermodynamic properties quite well, though, it gives less satisfactory results in the expanded liquid metals at low densities. Therefore, accuracy of theoretical studies on the thermodynamic properties of liquid metals is subjected to the accuracy of pair potential function describing the intermolecular interaction of these metals.

In this study, we present density dependent equations of state for metal, nonmetal, and metal-nonmetal transition states of fluid mercury. For transition state, density dependent potential functions are applied and successfully determine an equation of state which shows inflection from metal to nonmetal. The density range of the transition state is characterized and the method for accurate determination of molecular parameters of the potential function is investigated.

## 2. Density dependence potential model

The structure factors of liquid mercury were measured by Tamura and Hosokawa,[14] using x-ray along liquid-vapor coexisting line. Using these structure factors and performing inversion method, potential functions were extracted by Munejiri *et al*. in three states: metallic state (1273 K, 10.98 gr/cm$^3$), metal-nonmetal transition state (1673 K, 9.25 gr/cm$^3$), and in non-metallic state (1803 K, 6.8 gr/cm$^3$).[12] The characteristics feature of the effective pair potential function $\phi(r)$ thus obtained are as follows: **a**) the attractive well of $\phi(r)$ for metallic state is broader than those of non-metallic state; **b**) $\phi(r)$ of the nonmetallic state is similar to the Lennard-Jones (12-6) potential function; **c**) the repulsive part of $\phi(r)$ is harder than those of Rubidium and Cesium; **d**) when the state of liquid mercury changes from metallic to the metal-nonmetallic transition state, the repulsive part shifts to shorter distance and shifts back as the non-metallic state is approach.

The above three states have been paid attention also by Okurmura and Yanezawa,[15] who have made progress to describe the bulk viscosity of density dependent potential systems. Bulk viscosity is a measure of dilatational distortion of a liquid and is important in the case of compressible liquid.[16,17,18] Since the metal-nonmetal transition does not occur at definite density, the density dependent potential function $u(r)$ at all densities becomes the potential function of intermediate density asymptotically. On the other hand, the transition state comprises species interacting either on a metallic path or on a nonmetallic one. Therefore, over the whole range of density of fluid mercury

$$u(r) = f(\rho)u_1(r) + (1 - f(\rho))u_2(r) \quad (1)$$

where $f(\rho)$ is applied to specify extend of transition from state with potential function $u_1(r)$ to the state with potential function $u_2(r)$.[15] For liquid mercury, metal-nonmetal transition occurs at $\rho \approx 8 - 9\,g/cm^3$. Therefore $f(\rho)$ has the form

$$f(\rho) = \begin{cases} 1 & \rho \ge 11\,g/cm^3 \\ (\rho - 8)/3 & (8 < \rho < 11)\,g/cm^3 \\ 0 & \rho < 8\,g/cm^3 \end{cases} \quad (2)$$

## 3. Density Dependent Equation of State

To derive equation of state for mercury, we follow closely the pervious model for dense liquid cesium metals. The method is outlined as follows while the details can be found elsewhere.[19,20,21] First, the total interaction potential energy of N-atom-liquid system $U(r_1, r_2, \cdots, r_N)$ is approximated as the sum of pairwise interaction potential energy $u(r_{ij})$ of the atom *i* with atom *j* located at distance $r_{ij}$ in the *liquid* *state*. The summation is restricted to the nearest neighboring atoms and thus $U(r_1, r_2, \cdots, r_N) = (N/2)u(r)$, where r is the interatomic distance. [Here for the time being we do not include the coordination number. See section 4.] Then, the thermodynamic equation of state is solved for pressure of the system, where the internal energy involves total interaction potential energy and kinetic energy. Finally isotherms are obtained for the three states by application of pair potential specified by eqs. (1) and (2).

We apply Lennard-Jones (12-6) potential function to low density with nonmetallic character (e.g., the state with $u_2$). The polarizability of nonmetal mercury shall be appreciably large to follow an inverse power law $1/r^6$. For metal state specified by pair potential $u_1$, we apply LJ (15-9) potential function, which indicates the less attraction and a hard repulsion interactions characteristics of closed shell ground state of mercury. Both $u_1$ and $u_2$ have harder repulsive part than the pair potential of open shell heavy alkali metal in accord to the results obtained by inversion method using experimental pair correlation function.[12]

From the model outlined above, the isotherm thus obtained for metal state is in the form

$$(Z-1)V^3 = A\rho^2 + B \qquad (3)$$

where Z is the compression factor and V is the molar volume. Constants A and B depend on temperature and are a measure of repulsive and attractive parts of the potential function $u_1$, respectively. Equation (3) indicates that isotherms of $(Z-1)V^3$ are linear functions of $\rho^2$. Notice that for metal state a pure metallic interaction is assumed, where often pseudopotential is applied for first principle calculation.[14] The form of $f(\rho)$ restrains the application isotherm (3) to metal state for $\rho \geq 11\,g/cm^3$. From the low-density side of the domain [see eq. (2)], this isotherm tends to the isotherm with characteristic of mixed potential function $u_1$ and $u_2$. Determination of the parameters of the isotherm (3) at a given temperature leads to the calculation of molecular parameters of the corresponding metallic potential function:

$$\varepsilon_m = \frac{2RT}{5N_A}\left(\frac{B^5}{A^3}\right)^{1/2}, \qquad \sigma_m = K_{cell}\left(\frac{-3A}{5B}\right)^{1/6}, \qquad (r_{min})_m = (5/3)^{1/6}\sigma_m \tag{4}$$

where $\varepsilon$ is the potential well-depth, $N_A$ is the Avogadro's number, $\sigma$ is the hard-sphere diameter, the subscript m stands for metallic state, and RT has its usual meaning. $K_{cell}$ is a constant characteristic of unit cell and can be determined analytically if the unit cell assumed for the metal fluid system is known. In this investigation, its form consists of $K_{cell} = r_{min}/V^{1/3}$, where $r_{min}$ is interatomic distance at potential minimum.

For the nonmetal state, the linear isotherm is derived in the form

$$(Z-1)V^2 = C\rho^2 + D \tag{5}$$

where C and D are constants characteristics of repulsion and attraction of the potential function $u_2$, respectively. Equation (5) indicates that $\rho^2$ dependence of $(Z-1)V^2$ is linear with slope C and intercept D. Again, one would expect to calculate the molecular potential parameters for $u_2$ with reasonable accuracy by using the constants of isotherm (5):

$$\varepsilon_n = \frac{RTD^2}{2N_A C}, \qquad \sigma_n = K_{cell}\left(\frac{-C}{2D}\right)^{1/6}, \qquad (r_{min})_n = 2^{1/6}\sigma_n \tag{6}$$

where subscript n stands for nonmetal state. Contrary to the simple form of isotherms for metal and nonmetal states, the isotherm of metal-nonmetal transition states represent the complexities inherent to the transition state in the form

$$(Z-1)V^2 = E\rho^4 + F\rho^3 + G\rho^2 + H\rho + I \tag{7}$$

where E, F, G, H, and I, are constants characteristics of the molecular parameters of LJ (12-6) and LJ (15-9) potential functions.

The consistency of experimental data with the suggested potential function indicates that the effective interaction in the metal state is shorter in range than that of nonmetal state. This is probably due to the density effect. That is, in the dense metal state the correlation between atoms is high leading to a highly ordered system, whereas in the nonmetal state the correlation is poor and thus the electrostatic fields of charged species extent readily over several atomic diameters. Experimental data also state that as the temperature is increased the intermolecular distance does not change much but the coordination number decreases.[22] This is consistent with our finding based on the application of $u_1$ and $u_2$ which leads to isotherm (7) with

$$E = \frac{15}{2}\left(\frac{200.6}{3\times10^6}\right)\left(\frac{5}{3}\right)^{3/2}\frac{N_A\varepsilon_m}{RT}\left(\frac{\sigma_m}{K_{cell}}\right)^{15} \tag{8}$$

$$F = -\left(\frac{50}{3}\right)\left(\frac{5}{3}\right)^{3/2}\frac{N_A\varepsilon_m}{RT}\left(\frac{\sigma_m}{K_{cell}}\right)^{15} - 10\left(\frac{200.6}{3\times10^6}\right)\frac{N_A\varepsilon_n}{RT}\left(\frac{\sigma_n}{K_{cell}}\right)^{12} \tag{9}$$

$$G = -5\left(\frac{5}{3}\right)^{3/2}\left(\frac{200.6}{3\times10^6}\right)\frac{N_A\varepsilon_m}{RT}\left(\frac{\sigma_m}{K_{cell}}\right)^{9} + \left(\frac{88}{3}\right)\frac{N_A\varepsilon_n}{RT}\left(\frac{\sigma_n}{K_{cell}}\right)^{12} \tag{10}$$

$$H = 10\left(\frac{5}{3}\right)^{3/2}\frac{N_A\varepsilon_m}{RT}\left(\frac{\sigma_m}{K_{cell}}\right)^{9} + 6\left(\frac{200.6}{3\times10^6}\right)\frac{N_A\varepsilon_n}{RT}\left(\frac{\sigma_n}{K_{cell}}\right)^{6} \tag{11}$$

$$I = -\left(\frac{44}{3}\right)\frac{N_A\varepsilon_n}{RT}\left(\frac{\sigma_n}{K_{cell}}\right)^{6} \tag{12}$$

## 4. Results and discussion

The atomic vapor state of mercury can be described by the same type of interaction potential as simple fluids and a Lennard-Jones (12-6) potential function is applicable. This means the law of corresponding states is applicable to the atomic mercury vapor as well as simple fluids. However, the deviation from this behavior is seen as temperature is increased. The reason for these deviations is that the internal degrees of freedom of mercury atom can be excited, and in particular ionization occurs.[3] For an accurate theoretical treatment, an ionization equilibrium between Hg and $Hg^+$, $Hg^{2+}$, and $\bar{e}$ has been assumed, and association processes leading to $Hg_2$ and molecular ions $Hg_2^+$ have been included.[3] Therefore, for the details of metal-nonmetal transition in a more rigorous microscopic treatment of mercury fluid, the change of internal degrees of freedom must be considered. Then the interaction potential itself becomes dependent on density and temperature.

Repulsive branches of effective ion-ion potential of mercury and LJ (m-n) potential with m>12 are almost identical with hard sphere potential, and thus independent of density. This feature actually leads to the application of hard sphere potential for most practical purposes.[3] The interionic potential at freezing temperature is purely repulsive around nearest-neighbor distance. The attractive branch changes from that of a screened coulomb interaction in liquid to an induced dipole-dipole interaction characteristic of nonmetal in atomic vapor state.

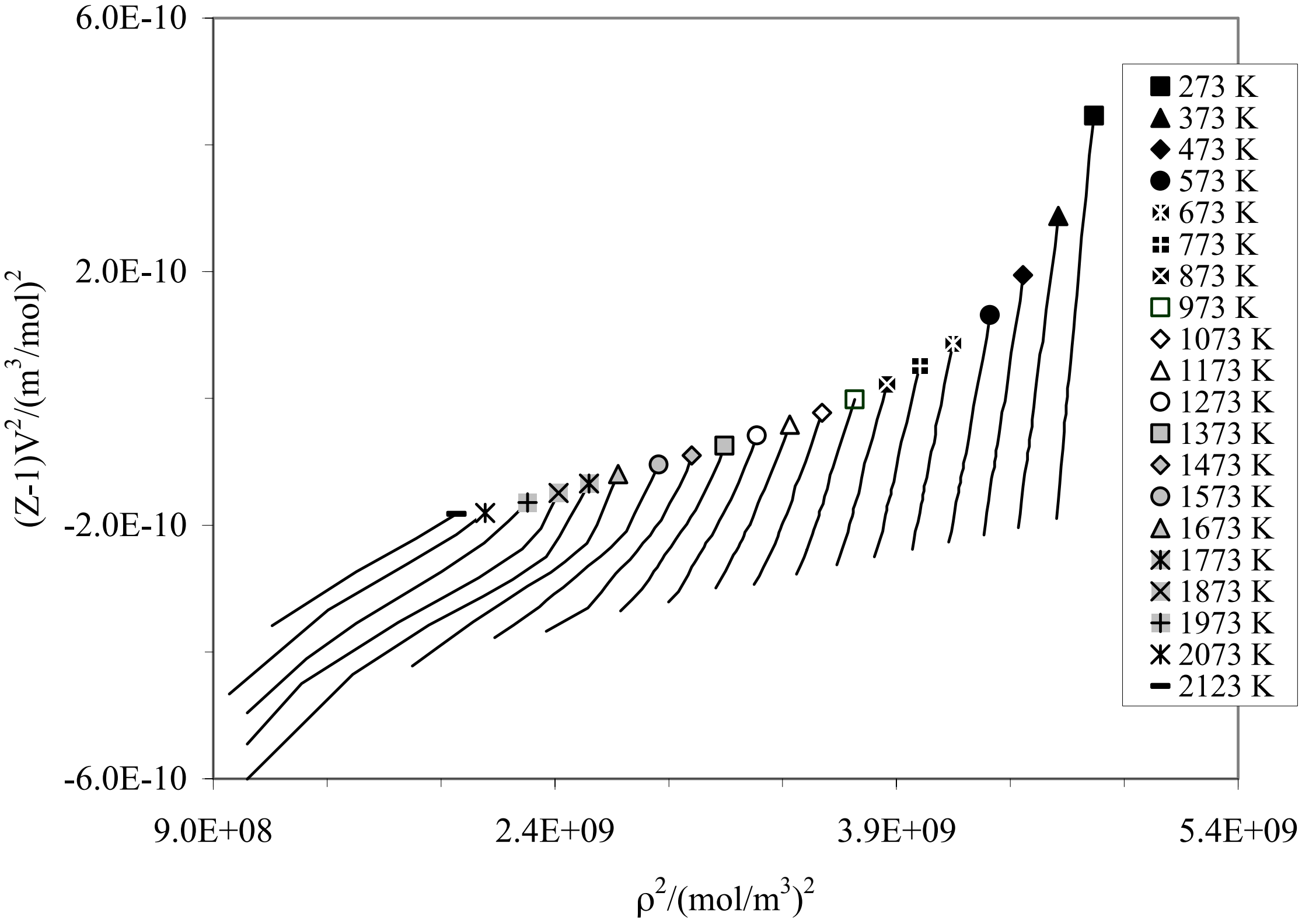

**Figure 1**. Application of the isotherm of non-polar, nonmetal fluids to compressed fluid mercury over metal, nonmetal, and metal-nonmetal transition states. The deviation from linear behavior is enhanced most in the transition state.

In a preliminary consideration for evaluation, we have applied to liquid mercury LJ (12-6)- and LJ (6-3)-potential functions, which have been shown to be applicable to normal liquids and liquid cesium metal, respectively. The resulted isotherms are shown in Figures 1 and 2 for metal, nonmetal, and transition states. The experimental $P\rho T$ tabulations measured for mercury have been used.[23] [The selected data from available data are in the pressure range 200-5000 atm at T=273.15 K, and to 3500-5000 atm at T=2123.15 K; the data used for other isotherms given below are selected so as to follow the constraint for $f(\rho)$ in (2).] Both potential functions result in isotherms that are linear from 273.15 K up to 1173.15 K. This temperature range corresponds to the (available) density $13.85 - 11.18\ g/cm^3$ of the compressed liquid mercury. It is concluded that for fluid densities larger than $11\,g/cm^3$ both isotherm are perfectly linear. In the same way, the isotherms tend to a linear behavior as the critical point is approached and the liquid-vapor phase transition is occurred. As we see later this corresponds to the liquid with $\rho < 8\,g/cm^3$. However, both isotherms markedly deviate from linear behavior in the metal-nonmetal transition region, indicating that a particular force law governs the thermodynamic properties. The properties of liquid mercury such as electrical conductivity indicate metal properties in the former range and nonmetal properties in the later range. Simultaneous measurement of density and electrical conductivity of compressed liquid mercury have been performed by Yao and Endo.[24] They have presented results of their measurements by graphs. The result are limited to transition state (e.g., density range 8-12 gr/cm$^3$ corresponding to the temperature range 1273-1773 K and up to 2200 atm). Electrical conductivity shows a singularity at 1753 K corresponding to the density in the range 9 to 8 gr/cm$^3$.

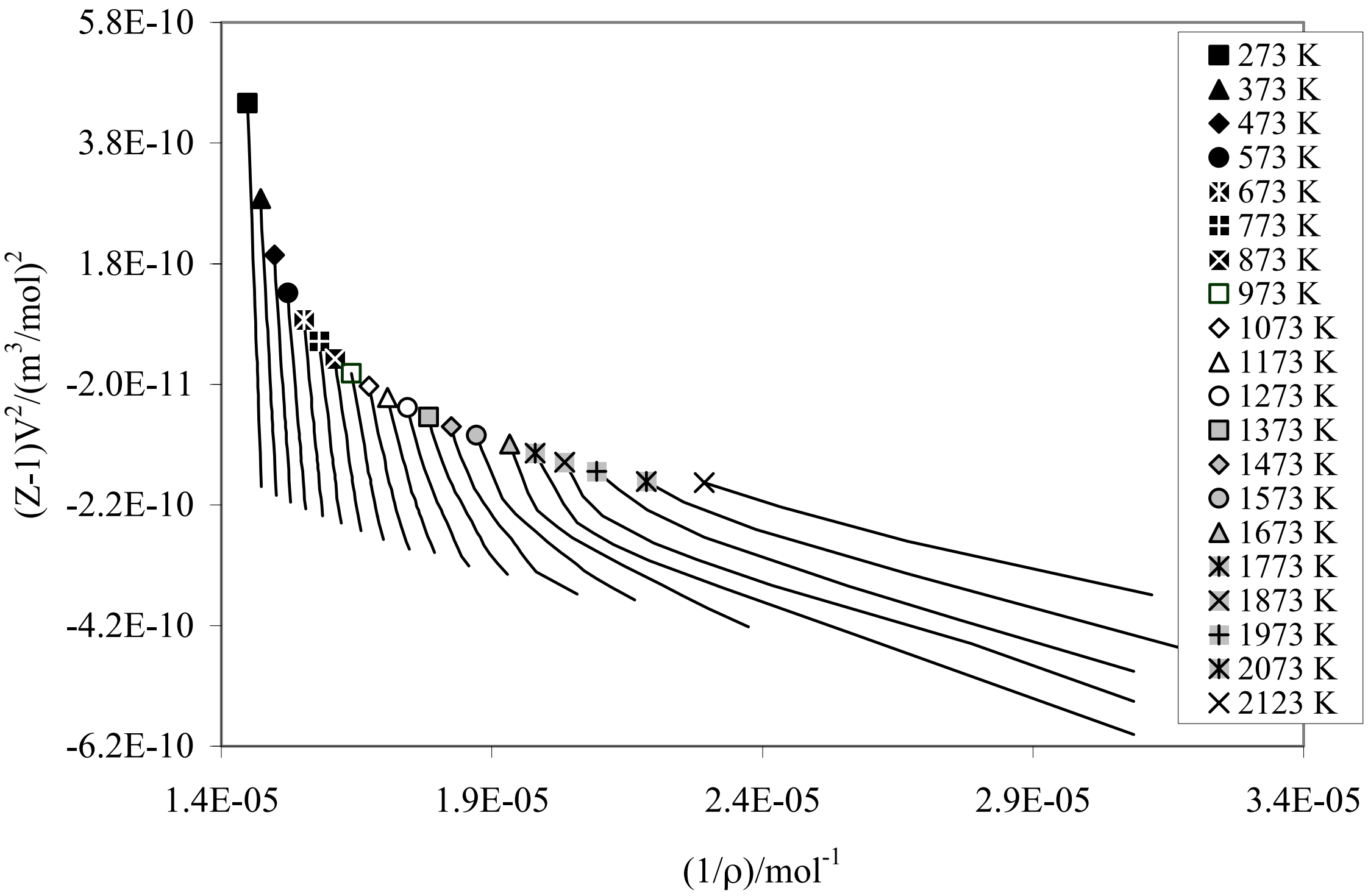

**Figure 2**. Application of the isotherm of cesium metal fluid to compressed fluid mercury over metal, nonmetal, and metal-nonmetal transition states. The deviation from linear behavior is enhanced most in the transition state. The range of linear behavior in the nonmetal state is more enhanced than in figure 1.

Below this temperature, the electrical conductivity represents a certain regime which can be attributed to a metallic state, but above this temperature not enough data given to allow making a clear conclusion about the nature of (nonmetallic) state. The same singularity and behavior are seen from pressure variation of density.[24] Therefore, as long as the fluid system consists of pure identical particles interacting with the same force law, the same power law strictly governs the thermodynamic properties.[25]

The isotherms of eq. (3), which have been resulted by using LJ (15-9) as the effective pair potential in metal state, are shown in Figure 3. The reason for selecting such a potential function with hard repulsion is the closed shell $6s^2$ electronic structure. In other words, the major interaction in liquid mercury (in metal state) is due to screened ion-ion, which can be described by a hard repulsive branch. Indeed we have applied LJ (15-9) to produce an accurate equation of state for metal state, and additionally to yield an analytical equation of state, which could fit suitably the experimental PρT data in metal-nonmetal transition region. (See next paragraph.)

The isotherms of eq. (5) for which supercritical $P\rho T$ data of compressed mercury have been used are shown in Figure 4. Our considerations show that the linear behavior persists well if, at a given temperature, only data with $(5.8 < \rho \leq 8)\,g/cm^3$ are included, where the lower limit is the critical density $\rho_c$. This density limit is rather sharp so much the linear behavior is deteriorated extensively, otherwise. This indicates that for $\rho \leq 8\,g/cm^3$ the fluid mainly consist of neutral atoms with induced dipole-dipole interaction. Our finding is consistent with the literature report in which the non-metal state is described well by LJ (12-6) potential function.[3,12]

The success of the present work can be attributed to the derivation of the fourth-order eq. (7) as the equation of state for metal-nonmetal transition range (See Figure 5.). From the available experimental data, we have chosen the temperature range (1473.15-1848.15 K) which includes low-density metal state and high-density nonmetal state to test this equation of state. In the insert of Figure 5, the density variation of $(Z-1)V^2$ shows an enhancement of inflection point from metal branch to nonmetal one. In particular, at 1748.15 K there is a marked inflection point, though all isotherms show such a behavior slightly.

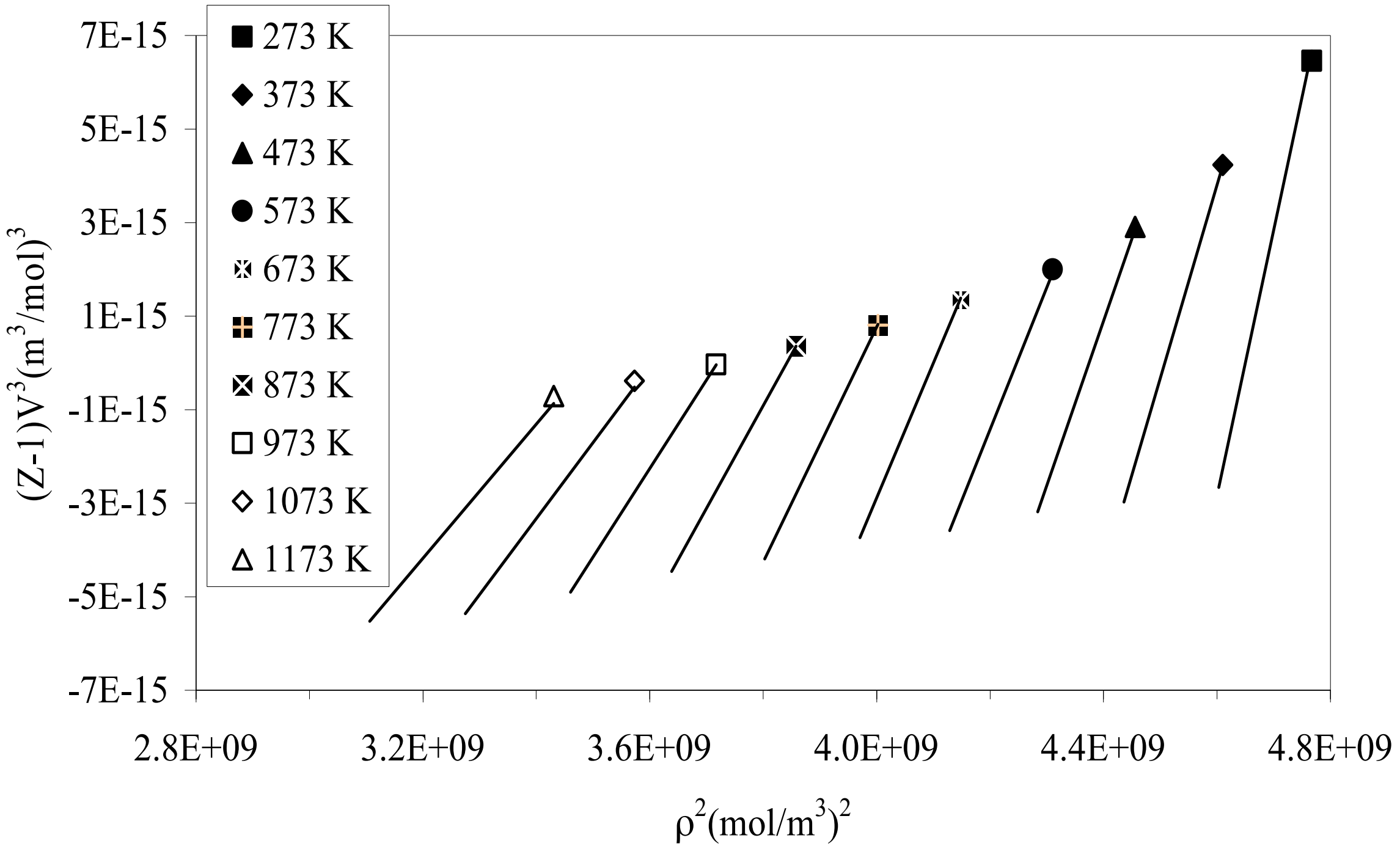

**Figure 3**. Plots of isotherms eq. (3) in the density range where the metal behavior is enhanced. The plots are subjected to the restriction of f(ρ) in (2).

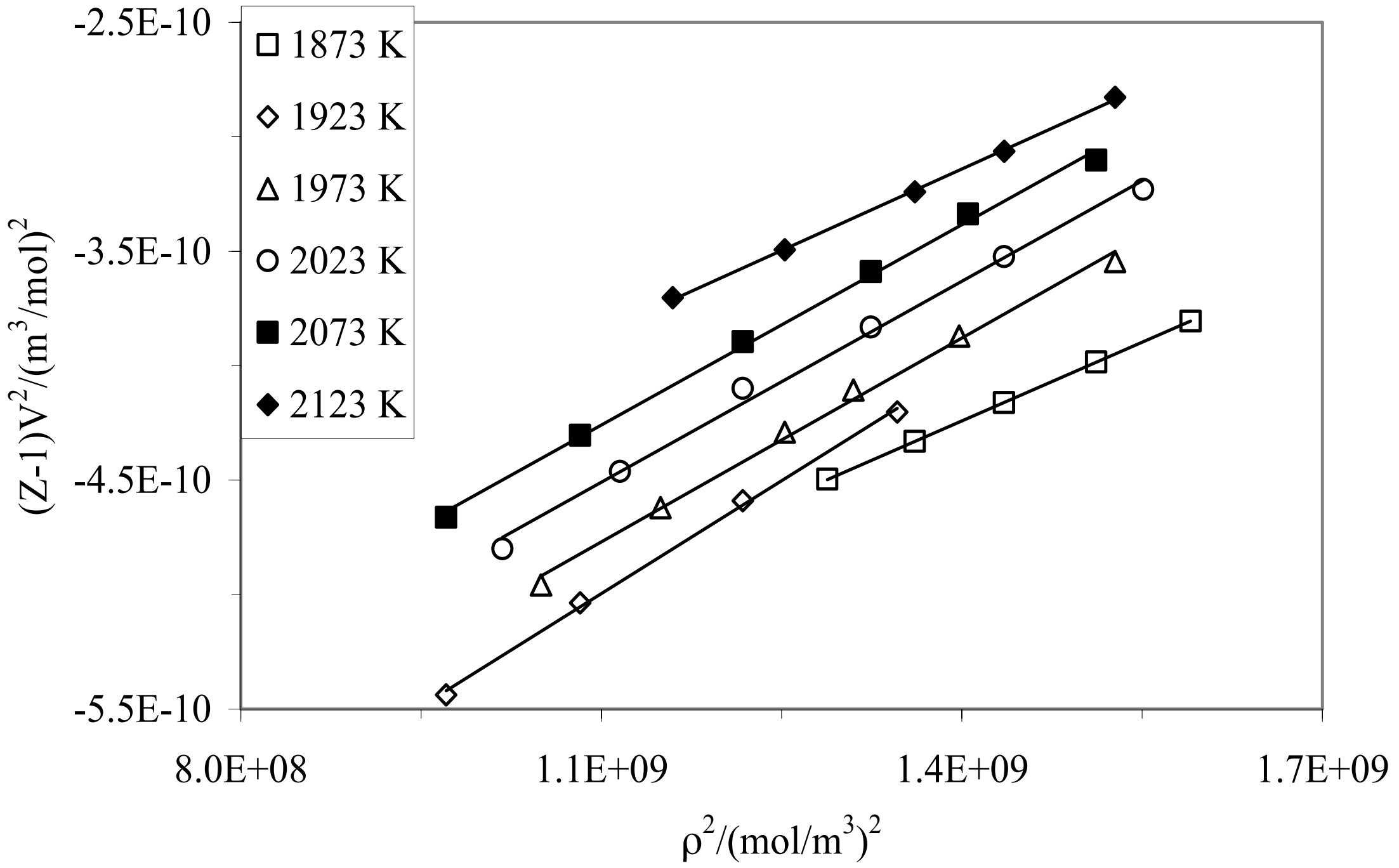

**Figure 4**. Plots of isotherms of eq. (5) for nonmetal state. The plots subjected to the restriction of f(ρ) in (2).

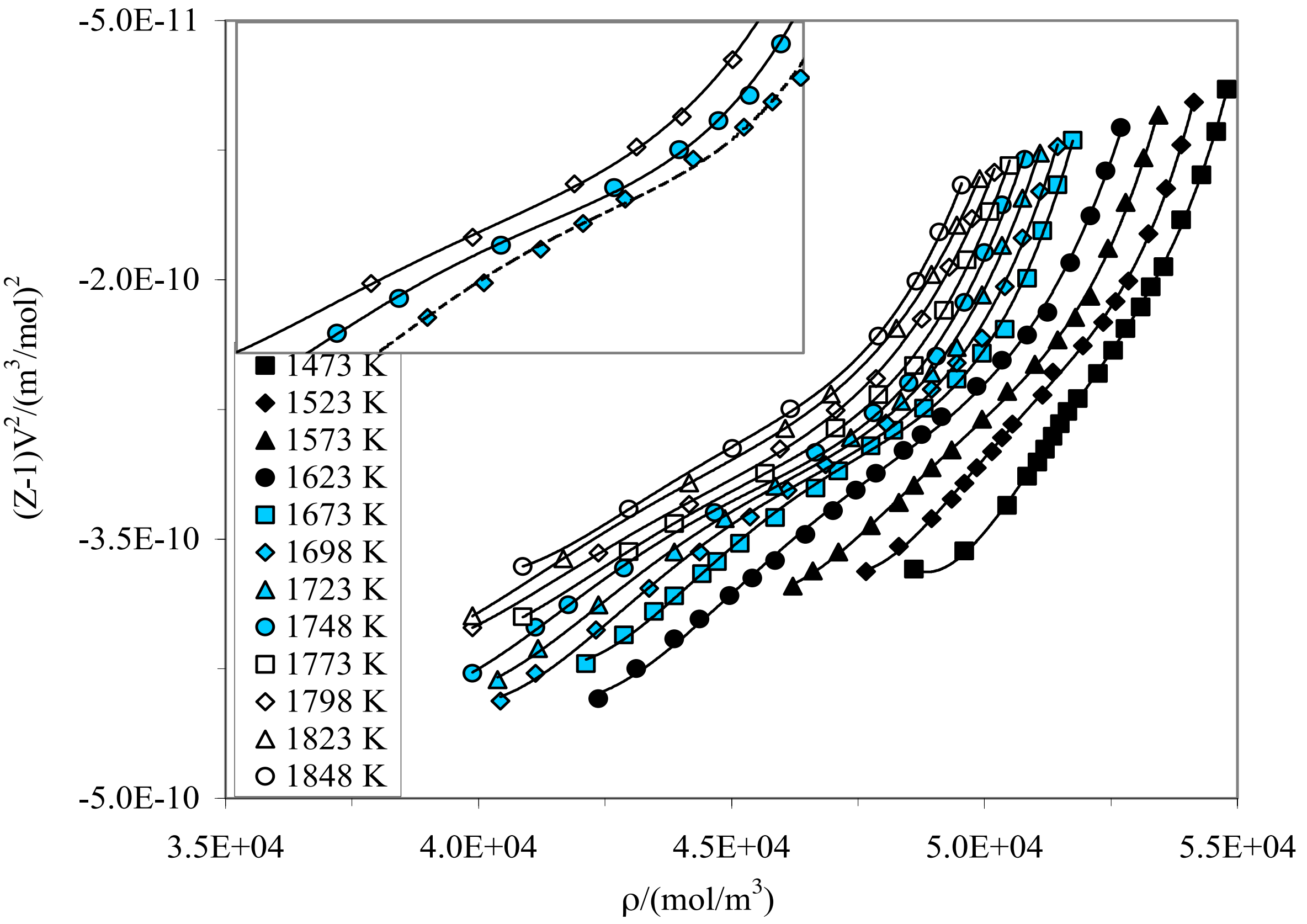

**Figure 5**. Plots of isotherms eq. (7) for metal-nonmetal transition range. The insert are enhancements of isotherms at 1698 K, 1748, and 1798 K.

The value of $r_{min}$ calculated for metal, nonmetal, and metal-nonmetal transition regions are shown in Figure 6. We have used values of $r_{min}$ from the plots of pair correlation function versus interatomic distance,[2] determined at saturation pressures, and have calculated $K_{cell}$. We have noticed that $K_{cell}$ remains almost constant over the whole liquid range. This may be attributed to the fact that $K_{cell}$ is proportionality constant between two characteristics parameters of the fluid system, and thus we have applied this value, in spite of the fact we are confronting with the problem of applying a saturation property to solve for a properties at compression, in all calculations. The typical calculated value of $r_{min}$ at 273.13 K is 3.00 Å, which is in excellent agreement with experimental value of 3.01 Å (within 0.33%).[26] Thus, the method of this study reproduces the molecular potential parameter $r_{min}$ reasonably accurate. The experimental value of $r_{min}$ (and ε/k as well) have been determined by integration using available experimental second virial coefficient of mercury and LJ (12-6) potential function.[26] It is noticeable that $r_{min}$ (in this study) increases slightly with temperature (by 5.66%) at T=1373.15 K (at which the transition is admitted).

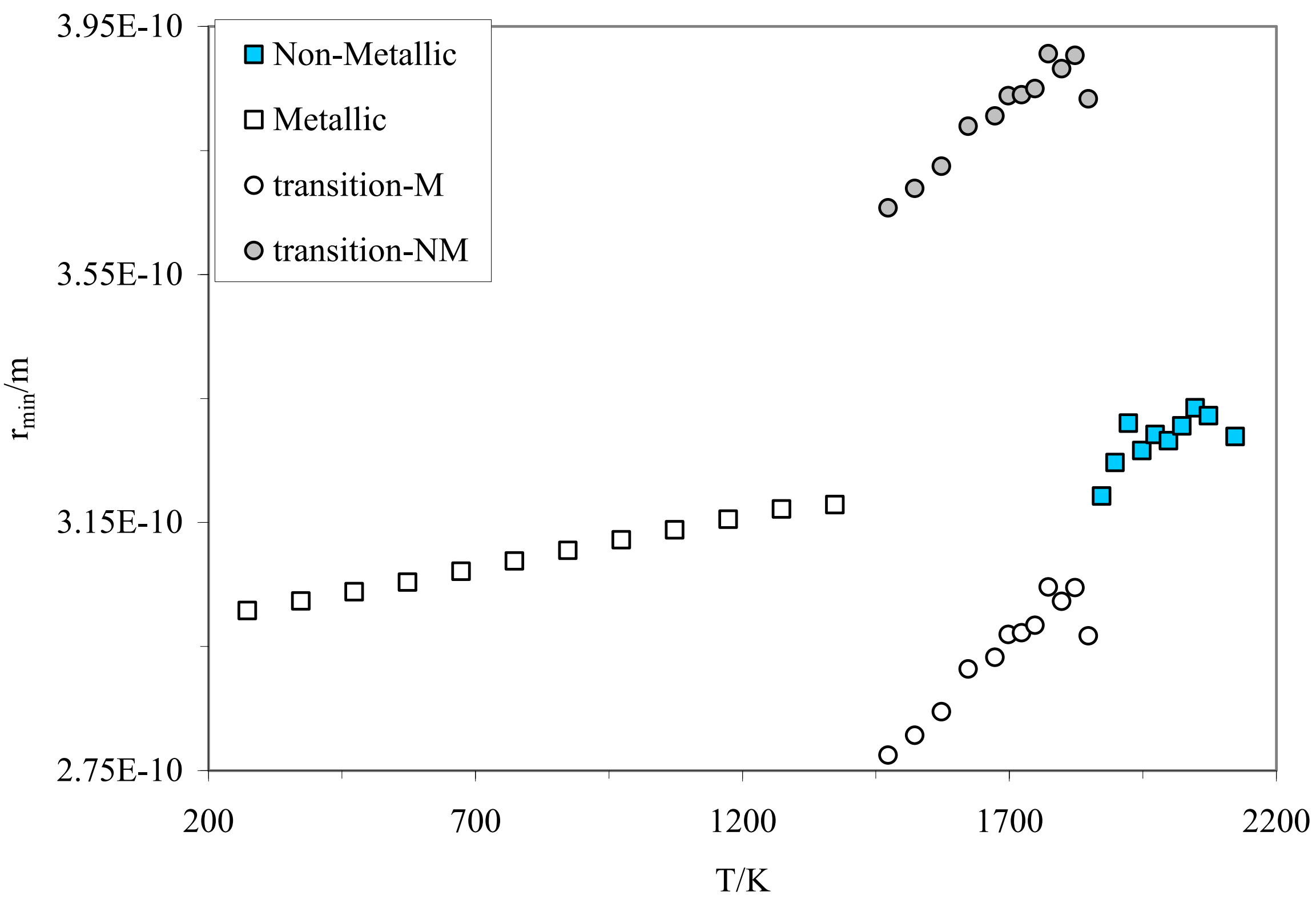

**Figure 6**. Plots of $r_{min}$ versus T for metal, nonmetal and transition states. The two plots in the transition region are resulted from solution of the relations (8)-(12).

The plots of $(\varepsilon/k)$ versus temperature (with k being the Boltzman constant) for metal and nonmetal ranges are shown in Figure 7. These values of $(\varepsilon/k)$ are calculated disregarding the coordination number of liquid mercury. For metal state, we have calculated $(\varepsilon_m/k)$ by using the Eq. (4) along with isotherm (3). The coordination number of mercury has been determined experimentally at saturation pressures, [27] and typically equals to 9.36 at $13.55\,g/cm^3$ corresponding to $T = 273.15\,K$. Thus, for a pair of mercury atom at this temperature, $(\varepsilon_m/k)_{calc.}/(\varepsilon/k)_{exp\,t.} = 981/1030 = 0.952$, where $(\varepsilon/k)_{exp\,t.}$ is the reported molecular potential parameter of LJ (12-6) obtained by using second virial coefficient.[26] Notice that at $T = 273.15\,K$ the available $P\rho T$ data are in the range 200-5000 atm. On the other hand, at T=1873.15 K, PρT data are available in the pressure range 2200-5000 atm, however, the data in the range 2200-2600 atm have been applied to follow restrictions for f(ρ) in (2). This excludes also densities less than critical density $\rho_c = 5.8\,g/cm^3$. At this temperature $(\varepsilon_n/k)_{calc.}/(\varepsilon/k)_{calc.}$ $= 428.3\,K/1030\,K = 0.416$, for which the coordination number equals 5.31.[23] It can be concluded that the method of this study can reproduce reasonably well the molecular parameters of pair potential function at low temperature, which for their calculations subcritical experimental data of compressed liquid mercury, comparable roughly with data at saturation pressures, are involved. [The critical point data are $T_c = 1751\,K$, $P_c = 1695\,atm$, and $\rho_c = 5.8\,g/cm^3$.] At the low temperature, the deviation of the above ratio from unity (e.g., 0.952) can be attributed to the fact that liquid mercury has been investigated at pressures range just somewhat higher than saturation pressures. On the other hand, at high temperature, the large deviation of $(\varepsilon_n/k)_{calc.}$ from that of an isolated pair can be attributed to the high pressures (supercritical) data, which pushes pair of mercury atom further inside the $r_{min}$ distance with more effective repulsive potential, thus lowering the potential well depth.

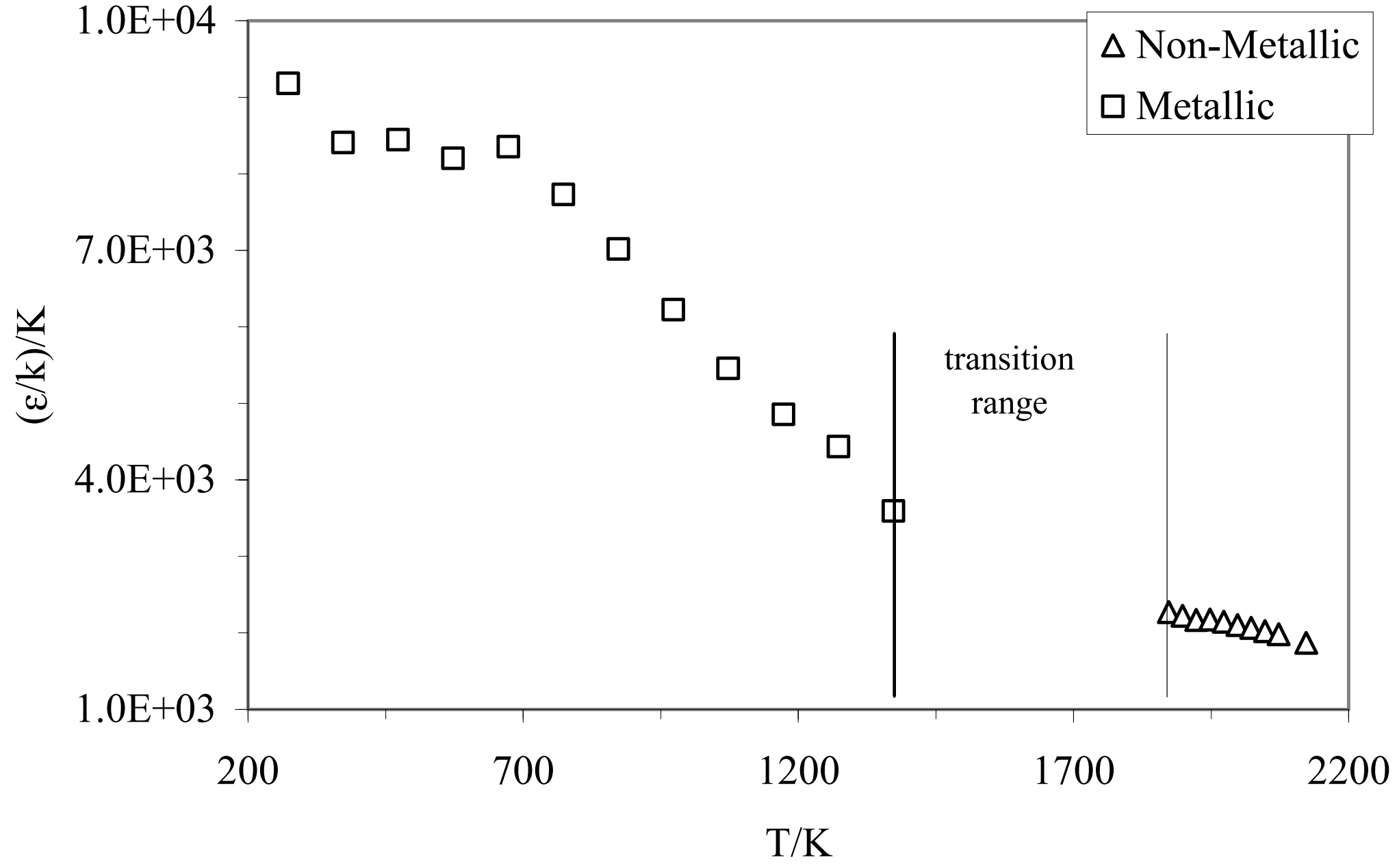

**Figure 7**. Plots of $(\varepsilon/k)$ versus T for metal and nonmetal states. No coordination numbers were applied.

Since the analytical form of $K_{cell}$ is not known, we have calculated its value by using reported experimental $r_{min}$,[2] in $r_{min} = K_{cell}V^{1/3}$ at $T = 273.15\,K$. We have noticed that $K_{cell}$ remains almost constant over range $T = 273.15 - 2000\,K$ and does not affect the values of $(\varepsilon/k)_{calc.}$

Figure 8 shows the plots of $(\varepsilon/k)$ versus temperature for the metal-nonmetal transition region obtained by using relations (8) to (12). Essentially simultaneous solution of (8)-(12) is appreciated for covering any uncertainty in the values of coefficient in Eq. (7), which may rise due to unsatisfactory fitting. However, we can not explain the large differences between $(\varepsilon/k)_{calc.}$ and $(\varepsilon/k)_{expt.}$ by 2 to 3 order of magnitudes with respect to that of metal region, and 1-2 order of magnitudes with respect to that of nonmetal region (See Figure 7.). Because $\varepsilon$ and $\sigma$ are dependent parameters in the transition region, the metallic (nonmetallic) value of $r_{min}(\propto \sigma)$ in this region is rather smaller (larger) than the corresponding value in the metallic (nonmetallic) region (See figure 6.). A clear explanation for the behavior of either parameter may lead to understanding of the other one.

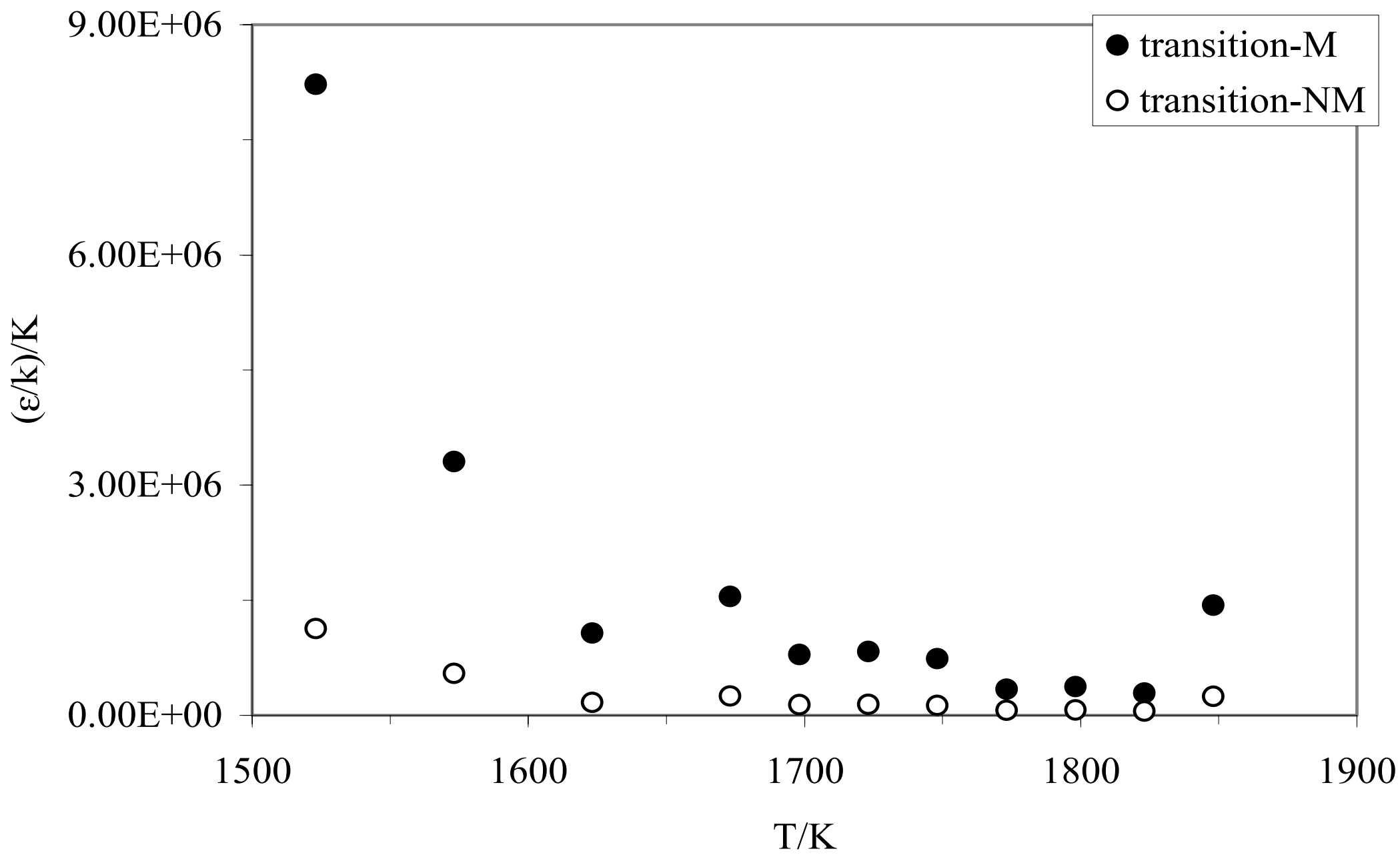

**Figure 8**. Plots of $(\varepsilon/k)$ versus T for the metal-nonmetal transition range. The two plots are for the transition region, and are resulted from simultaneous solution of the relations (8)-(12).

From metal side of the transition region, there exists a step-side rising edge which diminishes sharply as density decreases towards nonmetal side of the transition region. Despite lack of reasonable explanation, yet this is a singularity characteristic of a phase transition. Interestingly, the experimental bulk viscosity of mercury increases when density decreases to the density of nonmetal region, and passes through a maximum, almost at the middle of transition region corresponding to $9.5-9.75\,\mathrm{g/cm^3}$.[15] The simulation for bulk viscosity using pseudopotential method and the restriction for f(ρ) results in the same singularity though the rising edge is towards low-density side of the transition region.

## 5. Conclusion

Interatomic interaction potential energy functions have been used to derive equations of state in metal, nonmetal, and metal-nonmetal transition states of compressed fluid mercury. Density-dependent potential functions have been used, specifically for the metal-nonmetal transition range of fluid mercury, which has been shown to be limited to density range $(8<\rho<11)\,\mathrm{g/cm^3}$. The equations of state for metal and nonmetal states have simple forms, and the complexities associated with the metal-nonmetal transition have led to a rather complex fourth-order equation of state. The values of potential well-depth ε in metal region are in agreement with experimental values typically within 5.0% of an isolated pair of mercury-atom. However, calculated ε values of nonmetal found to be much smaller than a pair of isolated atom, attributing exceedingly large repulsion due to high pressure $P\rho T$ data used, which pushes atoms further together inside the $r_{min}$ distance. The calculated values of ε for transition region show singular behavior characteristics of a phase transition, though we can not explain a higher 2-3 order of magnitudes values with respect to metal state. The value of $r_{min}$ for metal state is in agreement with that of isolated pair within 0.33%. Also in the transition region $r_{min}$ calculated for nonmetal is higher (but finite value), and at the rising edge it increases by about 15% with respect to metal state. Singularities in the transition range in this study have shown enhancement on energy scale rather than on the structure scale.

**Acknowledgement**

The authors are indebted to the research council of Shiraz University for supporting this study. Authors are also grateful to H. Okumura of the Keio University for providing preprint (Ref. 15) of their studies prior to publication.